%% file: _2021_ArXiv-Francisco_Stefano-Error_Mitigation_using_Classical_Codes.tex
\documentclass[a4paper,11pt]{article}
\usepackage[utf8]{inputenc}
\usepackage[T1]{fontenc}
\usepackage[UKenglish]{isodate}
\usepackage[margin=2cm]{geometry}

\usepackage{cite}
\usepackage{amsmath,amssymb,amsfonts}
\usepackage{algorithmic}
\usepackage{graphicx}
\usepackage{textcomp}
\def\BibTeX{{\rm B\kern-.05em{\sc i\kern-.025em b}\kern-.08em
    T\kern-.1667em\lower.7ex\hbox{E}\kern-.125emX}}

\usepackage{bbm} 
\usepackage{amsmath,amssymb,amsfonts,amsthm} 

\usepackage{authblk}

\usepackage{multirow}

\usepackage{verbatim}

\usepackage{color, colortbl}
\usepackage{multirow}
\usepackage{easybmat}
\usepackage{physics}
\usepackage{epstopdf}

\usepackage{caption} 
\usepackage{subcaption}

\usepackage{widetext}

\usepackage{calc} 
\usepackage{adjustbox}

\usepackage{tikzit}
\input{sample.tikzstyles}

\providecommand{\keywords}[1]{\textbf{\textit{Keywords:}} #1}

\newtheorem{theorem}{Theorem}

\newtheorem{definition}[theorem]{Definition}

\begin{document}
\title{\vspace{-1.3cm}Error Probability Mitigation in Quantum Reading using Classical Codes}
\author[1,2]{Francisco Revson F. Pereira\footnote{revson.ee@gmail.com}}
\author[1,2]{Stefano Mancini}
\affil[1]{School of Science and Technology, University of Camerino, I-62032 Camerino, Italy}
\affil[2]{INFN, Sezione di Perugia, I-06123 Perugia, Italy\vspace{-1cm}}

\date{}

%
\maketitle              

\begin{abstract}
A general framework describing the statistical discrimination of an ensemble of quantum channels is 
given by the name of quantum reading. Several tools can be applied in quantum reading to reduce the error 
probability in distinguishing the ensemble of channels. Classical and quantum codes can be envisioned for this goal. 
The aim of this paper is to present a simple but fruitful protocol for this task using classical error-correcting codes. 
Three families of codes are considered: Reed-Solomon codes, BCH codes, and Reed-Muller codes. 
In conjunction to the use of codes, we also analyze the role of the receiver. 
In particular, heterodyne and Dolinar receivers are taken in consideration. 
The encoding and measurement schemes are connected by the probing step. As probe we consider coherent states. 
In such simple manner, interesting results are obtained.
As we show, for any fixed rate and code, there is a threshold under which using codes surpass optimal 
and sophisticated schemes. However, there are codes and receiver schemes giving lower thresholds. 
BCH codes in conjunction with Dolinar receiver turn out to be the optimal strategy for error mitigation in the quantum reading task.
\end{abstract}

\newline



\section{Introduction}
\label{sec:Introduction}
\input{./introduction}

\section{Preliminaries}
\label{sec:Preliminaries}
\input{./preliminaries}

\section{Proposed Protocol}
\label{sec:protocol}
\input{./description}

\section{Analysis of Error Probability using Classical Codes}
\label{sec:errorMitigation}
\input{./errorMitigation}

\section{Final Remarks}
\label{sec:Conclusion}
\input{./conclusion}

\section{Acknowledgments}
The authors acknowledge the financial support of the Future and Emerging Technologies (FET)
programme, within the Horizon-2020 Programme of the European Commission, under the FET-Open 
grant agreement QUARTET, number 862644.

\bibliographystyle{IEEEtran}
\bibliography{ref}

\end{document}

%% file: sample.tikzstyles
\tikzstyle{Red}=[fill=red, draw=black, shape=circle]

\tikzstyle{edge right arrow}=[->]

%% file: introduction.tex
\noindent

Quantum state discrimination composes an important part of several quantum computing protocols~\cite{Hirvensalo2003}. 
Quantum communication relies in the ability of the receiver to distinguish between a set of information carriers~\cite{Cariolaro2015,AndreasWinter2020}. 
The security of quantum key distribution protocols is based 
on the impossibility of perfectly distinguishing non-orthogonal states~\cite{Hirvensalo2003}. In both situations, where one needs to show that states are 
distinguishable or not, the set of quantum states are fixed and an analysis over the minimum achievable error probability is performed.
In order to decrease the error probability, the only possible method is optimizing the measurement apparatus. 
There are paradigms giving more freedom and, therefore, increasing the complexity for analyzing them.

A natural extension of the task of discriminating quantum states is envisioned in quantum channel 
discrimination~\cite{Childs2000,Acin2001,Gilchrist2005,Sacchi2005,Sacchi2005a,Wang2006,Duan2009,Hayashi2009,Harrow2010,Rexiti2021}, 
and, more generally, quantum reading~\cite{Pirandola2011}. Quantum reading considers the use of input and output quantum resources 
to enhance the retrieval of classical information stored in quantum channels. Actually, it considers that one is able to record 
bits of information in memory cells by storing a quantum channel picked from a given ensemble. The goal in quantum reading is to optimize the probing strategy 
as well as the encoding and decoding protocols to reduce the error probability in the discrimination process.

Efficient paths to quantum reading can be foreseen by the use of coding techniques~\cite{Pirandola2011a,Das2019,Pereira2020}. 
On the first hand, quantum error-correcting codes gives interesting candidates to probe the memory cells. The structure of their Hilbert subspace 
makes them reliable to some dissipator noise due to system-environment interaction~\cite{Pereira2021}. On other hand, 
the encoding process can be implemented using classical error-correcting codes in order to add redundancy and increase error mitigation. 
Focusing on error probability mitigation as figure of merit and using short length classical codes, we consider three families: Reed-Solomon codes, BCH codes, 
and Reed-Muller codes~\cite{Pellikaan2017}. These families have a large diversity of parameters and their values can be easily controllable. 
Additionally, the existing low-complexity encoding and decoding algorithms make the protocol proposed in this paper realizable with the current technology.

Following the previous reasoning, we consider quantum reading task where the ensemble is composed by two quantum channels. 
Each quantum channel is modeled by a pure-lossy bosonic channel with different transmissivities. 
For the analysis, we will consider the three families of codes mentioned before, two types of receivers, and a probing state. 
The first receiver is a heterodyne receiver~\cite{Cariolaro2015}. 
The use of heterodyne receiver is justifiable due to the phase-insensitivity property of pure-lossy channels. The second one is a Dolinar receiver~\cite{Cariolaro2015}. 
This is an adaptive receiver that can achieve the Helstrom bound on distinguishing two quantum states~\cite{Cariolaro2015}. There is an intrinsic complexity in implementing 
this receiver, mostly due to the adaptive and fast response characteristics. However, it has been implemented in practice, where its optimality was shown~\cite{Cook2007}.
Lastly, for the probing state, we consider coherent states. Therefore, we aim to show that the improvements and results obtained in the proposed scheme 
is due to the classical codes and receivers considered.

For any of the three families of classical error-correcting codes considered, we show improvements when compared with optimal strategies using coherent or squeezed states.
For fixed code, rate and receiver, we compute the average number of photons needed to surpass the optimal strategies. This value defines a threshold. There are 
strategies giving lower threshold. Using the heterodyne receiver, the best strategy is the BCH codes. However, there are some values of rate where Reed-Muller codes 
have similar performance. Using the Dolinar receiver, BCH and Reed-Solomon codes have almost the same performance. For lower values of rate, the BCH codes perform 
better than Reed-Solomon. The situation changes when the rate is above $0.45$. For any of the codes, the best performances are obtained using the Dolinar receiver.
We are able to achieve thresholds in this strategies for average number of photons below $18$. Furthermore, in order to achieve the threshold for a large range of 
rates, one does not need more than $6$ photons per probing state. 


This paper establishes important basis for the use of classical codes in quantum reading. In particular, it has taken the novel path of proposing an explicit use 
of short-length classical codes and show how much they impact in the error mitigation. The only related work in the literature is Ref.~\cite{Pereira2020}, 
where the authors proposed a polar coding and decoding schemes to achieve the reading capacity. 
However, the results in Ref.~\cite{Pereira2020} can only be appreciated for large code lengths. One can also find unrelated approach to quantum reading, 
such as Ref.~\cite{Banchi2020}, where it is shown that entanglement-assisted probing outperforms classical strategies on barcodes data, 
without any explicit analysis of encoding and decoding schemes.

This paper is organized as follows. In Section~\ref{sec:Preliminaries} we present the main concepts used through the paper. 
A description of the three families of classical codes is given. Additionally, we explain the task of quantum reading and the 
quantum channel model considered. Next, in Section~\ref{sec:protocol} we describe the proposed protocol. It is divided in 
three parts: probing strategy, encoding scheme, and decoding scheme. The performance analysis of the protocol is given in Section~\ref{sec:errorMitigation}. 
Lastly, we draw our conclusions in Section~\ref{sec:Conclusion}.

%% file: preliminaries.tex
This section introduces the main concepts of classical codes and quantum reading needed to this paper. 
We begin with a brief overview of cyclic codes and then specialize to Reed-Solomon and BCH codes. 
Subsequently, we show a construction method for Reed-Muller codes that is similar to Reed-Solomon codes.
Lastly, quantum reading problematic is introduced. The general concept is given, followed by a detailed description of 
the channel model adopted.

\subsection{Classical Codes}

\subsubsection{Cyclic Codes}
\label{sec:RScodes}

A linear code $C$ over $\mathbb{F}_q$ with parameters $[n,k,d]_q$ is called cyclic if for any codeword
$(c_0, c_1, \ldots, c_{n-1})\in C$ implies
$(c_{n-1}, c_0, c_1, \ldots, c_{n-2})\in C$. Defining a map from $\mathbb{F}_q^n$ to
$\mathbb{F}_q[x]/(x^n-1)$, which takes $\mathbf{c} = (c_0, c_1, \ldots, c_{n-1})\in \mathbb{F}_q^n$ to
$c(x) = c_0 + c_1 x + \cdots + c_{n-1}x^{n-1}\in \mathbb{F}_q[x]/(x^n-1)$, we can see that a linear code
$C$ is cyclic if and only if it corresponds to an ideal of the ring $\mathbb{F}_q[x]/(x^n-1)$.
Since that any ideal in $\mathbb{F}_q[x]/(x^n-1)$ is principal, then any cyclic code $C$ is generated
by a polynomial $g(x)$, which divides $(x^n-1)$. They are called generator polynomial.

A way to characterize the parameters of a cyclic code is by means of the generator polynomial and
its defining set. Roughly speaking, the defining set characterizes the common zeros of all polynomials $c(x)\in C$. 
More precisely, let $n$ and $q$ be relative prime, so $q^e \equiv 1 \mod n$, for some integer $e$. 
Fix an element $\beta$ of order $n$ in an extension $\mathbb{F}_{q^e}$ of $\mathbb{F}_q$.
We have that the defining set of $C$, which is denoted by $Z(C)$, is 
$Z(C) = \{i\in\mathbb{Z}_n\colon c(\beta^i) = 0\text{ for all }c(x)\in C\}$. The family of Reed-Solomon (RS) codes 
is a particular case of cyclic codes, where the generator polynomial has some
additional properties. 

\begin{definition}
            Let $\alpha$ be a primitive element of $\mathbb{F}_q$. Let $b\geq 0$, $n = q-1$, and $1\leq k\leq n$. 
            A cyclic code $RS_k(n,b)$ of length $n$ over $\mathbb{F}_q$ is a \emph{Reed-Solomon (RS) code} if 
            the generator polynomial is given by
            \begin{equation*}
                        g(x) = (x-\alpha^b)(x-\alpha^{b+1})\cdot\cdots\cdot(x-\alpha^{b+n-k-1}).
            \end{equation*}
            \label{Definition:RS}
\end{definition}

Since the minimal distance of any cyclic code is bounded from below by the maximum number of consecutive 
elements in $Z(C)$ and the Singleton bound says that the minimal distance of a $[n,k]$ code is not greater 
than $n-k+1$, we see that RS codes have minimal distance equal to to $n-k+1$. Thus, for fixed length and dimension, they have maximal minimal
distance possible and, therefore, they are named maximal distance separable (MDS) codes.

Even though the previous definition of RS codes describes them properly, there is a more practical way to construct RS codes.
Choose an enumeration $\mathcal{P} = (P_1,\ldots, P_n)$ of $n$ mutually distinct points in $\mathbb{F}_q$. 
Let $\mathbb{F}_q[X]$ be the set of all polynomials in the variable $X$ with coefficients in $\mathbb{F}_q$.
The RS code is given by
\begin{equation}
    RS_k(n,b) = \{ev_{\mathcal{P}}(f)\colon f\in\mathbb{F}_q[X], \text{deg}(f)< k\},
\end{equation}where $ev$ is the evaluation map defined by
\begin{eqnarray}
    ev_\mathcal{P}\colon \mathbb{F}_q[X] &\rightarrow&\mathbb{F}_q^n,\\
    f(X) &\mapsto& (f(P_1),\ldots,f(P_n)).
\end{eqnarray}

For the decoding scheme, we apply the Berlekamp–Massey algorithm~\cite[Section 5.4.2]{HuffmanVera:Book}. Since we are using small 
RS codes, meaning that the length of the RS codes is significantly shorter than LDPC or Turbo codes, the Berlekamp–Massey algorithmic fulfills 
our needs.

Before presenting BCH codes, we need to introduce the concept of minimal polynomials. Let $\beta\in\mathbb{F}_{q^e}$. 
The minimal polynomial over $\mathbb{F}_q$ of $\beta$ is the monic polynomial with smallest degree, $M(x)$, and coefficients
in $\mathbb{F}_q$ such that $M(\beta) = 0$. If $\beta = \alpha^i$ for some primitive $n$th root of unity $\alpha\in\mathbb{F}_{q^e}$, we denote 
the minimal polynomial of $\beta$ by $M^{(i)}(x)$.

\begin{definition}
    Let $\mathbb{F}_q$ be a finite field, $n$ and $q$ be relative prime, and $\alpha$ be a primitive $n$th root of unity. A cyclic code $BCH(\delta,b)$ 
    of length $n$ over $\mathbb{F}_q$ is a Bose-Chaudhuri-Hocquenghem (BCH) code of design distance $\delta$ if the generator polynomial is given by
    \begin{equation}
        g(x) = \text{lcm}\{M^{(b)}(x), M^{(b+1)}(x),\ldots,M^{(b+\delta-2)}(x)\},
    \end{equation}for some integer $b\geq 0$. If $n = q^e - 1$ then the BCH code is called primitive and if $b = 1$ it is called narrow-sense.
\end{definition}

The dimension of a BCH code is equal to $k=n-\text{deg}(g(x))$, similarly to the Reed-Solomon code case. The minimal distance can also be computed 
using the defining set of $BCH(\delta,b)$. However, there is no general formula for BCH codes. We would need to introduce the concept of 
$q$-ary cyclotomic coset modulo $n$ and analyze each code in order to obtain the respective dimension. Therefore, it is beyond the scope of this paper.
The performance analysis of the code is based on error probability and rate.

The encoding algorithm used for BCH codes is implemented via matrix multiplication, which has complexity $O(n^2)$. For the decoding algorithm,
the Berlekamp-Massey algorithm is also used.


\subsubsection{Reed-Muller Codes}
\label{RMcodes}

The construction of Reed-Muller codes is similar to the one presented for Reed-Solomon codes using evaluation map. 
The difference relies in the set of polynomials considered.
Take a vector space $\mathbb{F}_q^m$, where $m$ is an integer. Choose an enumeration $\mathcal{P} = (P_1,\ldots, P_n)$ of $n$ mutually distinct 
points in $\mathbb{F}_q^m$. The evaluation map is defined as 
\begin{eqnarray}
    ev_\mathcal{P}\colon \mathbb{F}_q[X_1,\ldots, X_m] &\rightarrow&\mathbb{F}_q^n,\\
    f(X_1,\ldots,X_m) &\mapsto& (f(P_1),\ldots,f(P_n)),
\end{eqnarray}where $\mathbb{F}_q[X_1,\ldots, X_m]$ is the set of all polynomials in the variables $X_1,\ldots,X_m$ with coefficients in $\mathbb{F}_q$.
We can define Reed-Muller codes by means of evaluating polynomials. 

\begin{definition}
    Let $\mathbb{F}_q$ be a finite field, and $r,m$ be integers such that $0\leq r < m(q-1)$. Let $n=q^m$ and 
    $\mathcal{P} = (P_1,\ldots, P_n)$ be an enumeration of all elements in $\mathbb{F}_q^m$. 
    A block code $RM(r,m)$ of length $n$ over $\mathbb{F}_q$ is a Reed-Muller code of order or 
    degree $r$ in $m$ variables if it is given by the set
    \begin{equation}
        RM(r,m) = \{ev_{\mathcal{P}}(f)\colon f\in\mathbb{F}_q[X_1,\ldots, X_m], \text{deg}(f)\leq r\}.
    \end{equation}
    \label{Def:RMcodes}
\end{definition}

It is possible to show that the dimension of a $RM(r,m)$ over $\mathbb{F}_q$ is equal to the size of the set~\cite[Proposition 5.4.7]{Pellikaan2017}
\begin{eqnarray}
    E_q(r,m) = \{\mathbf{e}\in\mathbb{N}_0^m\colon& 0\leq e_i\leq q-1 \text{ for all }i\nonumber\\
                                                  &\text{ and }e_1+\cdots+e_m\leq e\}.
\end{eqnarray}

The encoding algorithm used to construct the Reed-Muller codes is via generator matrix. We use the standard majority logic vote method due 
to Irving S. Reed \cite{Reed1954} in order to decode the received string.

\

\subsection{Quantum Memory Cell and Quantum Reading}\
\label{sec:quantum_memory}

We now provide a description of quantum memory cell and quantum reading suitable for this paper. A \textit{quantum memory cell} is defined as the set 
$\Phi = \{\mathcal{W}^x, p_x\}_{x\in\mathcal{X}}$ of quantum channels. For a fixed $x$, quantum input and output Hilbert spaces 
$\mathcal{H}_{B'}$ and $\mathcal{H}_{B}$, respectively, we have

\begin{eqnarray}
\mathcal{W}^x\colon \mathcal{D}(\mathcal{H}_{B'})&\rightarrow \mathcal{D}(\mathcal{H}_B)\\
\rho&\mapsto \mathcal{W}^x(\rho),
\end{eqnarray}where $\mathcal{D}(\mathcal{H}_{B'}),\mathcal{D}(\mathcal{H}_B)$ are the sets of input and output 
density states of the quantum channel $\mathcal{W}^x$, and $p_x = P_X\{X = x\}$ is the probability distribution law of $X$. 
We call $x\in\mathcal{X}$ the \textit{quantum memory cell index}. 
We consider $\cal{X}$ binary and the distribution of the random variable $X$ describing the label of the quantum channels to be 
Bernoulli with probability $p=1/2$. 

\begin{figure}[h!]
	\centering
	\includegraphics[width=.7\linewidth]{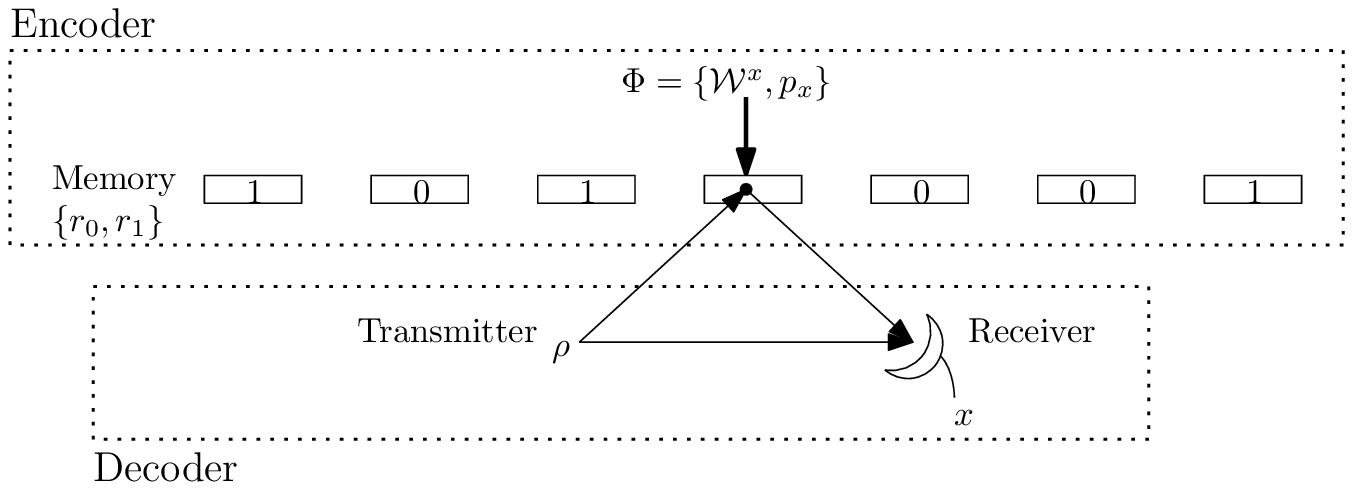}
	\caption{Quantum reading protocol.}
	\label{fig:quantum_reading}
\end{figure}

The quantum reading protocol consists in probing a memory cell in order to discriminate between its possible index values, i.e. between quantum channels.
Additionally, and more important to this paper, it is assumed that the encoder can use classical codes during the writing process on 
the quantum channels arising from the quantum memory cell in order to reduce the error probability. Let $\mathbf{c}$ be a codeword of 
a classical code. Then, the encoder is able to choose the 
quantum memory cell index, and, therefore, which channel to be placed in each position, according to $\mathbf{c}$. Since it is the encoder 
who chooses the quantum memory cell index, we can assume, without loss of generality, that a source code is performed on the information bits 
in order to produce indexes evenly distributed. A schematic of the quantum reading protocol is given in Fig.~\ref{fig:quantum_reading}.

For the quantum channel model, we use bosonic pure-lossy channels. They are a reasonable basic continuous variables 
model for optical memories. Thus, the binary channel ensemble is given by $\Phi = \{\mathcal{W}^x, p_x\}_{x=0,1}$, where $p_0 = p_1 = 1/2$, 
and $\mathcal{W}^x$ represents a pure-lossy channel with transmissivity $0\leq\kappa_x\leq 1$. The action of each $\mathcal{W}^x$ channel 
in the Heisenberg picture is described by the map
\begin{equation}
\hat{a}_B\rightarrow \sqrt{\kappa_x}\hat{a}_B - \sqrt{1-\kappa_x}\hat{a}_E,
\end{equation}where $\hat{a}_B$ is the annihilation operator of the probe mode and $\hat{a}_E$ is that of an environmental vacuum mode.
We are considering that the two parameters $\kappa_0$ and $\kappa_1$ assume values in the interval $[0,1]$, since it is considered that the optical memories are read by 
reflection.



\subsubsection{Optimal Error Probability}

In Ref.~\cite{Pirandola2011a}, it is analyzed the error probability under the optimization of the probe state and the measurement apparatus 
used in the receiver. The optimization considers that the decoder can probe the memory cells as many times as necessary and Helstrom's measurements 
are at disposal. Using coherent states, they show that the optimal error probability is
\begin{equation}
P_c = \frac{1 - \sqrt{1 - \exp[-\overline{n}(\sqrt{\kappa_0} - \sqrt{\kappa_1})]}}{2},
\label{ErrorProbCoherent}
\end{equation}where $\overline{n}$ is the average number of photons. On the other hand, they also consider 
nonclassical probes described via Einstein–Podolsky–Rosen transmitter. It is composed of $s$ pairs of signals and references, 
entangled via two-mode squeezing. The squeezing parameter $\xi$ of the two-mode squeezing state and the parameter $s$ are connected by 
the expression $\xi = \text{arcsinh}\sqrt{\frac{\overline{n}}{s}}$. Optimizing the error probability in terms of the parameter $s$, the 
following expression of the error probability is obtained
\begin{equation}
P_{s} = \frac{\exp(-\mu \overline{n})}{2},
\label{ErrorProbSqueezed}
\end{equation}where 
\begin{equation}
\mu = \frac{\kappa_0 + \kappa_1 + 2}{2} - 2\sqrt{\kappa_0 \kappa_1} - \sqrt{(1 - \kappa_0)(1 - \kappa_1)}.
\end{equation}In the following section we are going to analyze the error probability derived from the use of classical codes and compare 
it with the previous optimal error probabilities.

%% file: description.tex
This section is divided in two folds. Initially, we present the probing strategy applied to the memory cells. 
Afterward, the encoding and decoding protocols implemented to incorporate and retrieve information from the memory cells are shown. 
In particular, the heterodyne and Dolinar receivers used to measure the output probe state are described.

\subsection{Probing Strategy}
Choosing the probing state is an important step in any metrological system. However, our focus is on the improvement of using classical codes. 
Therefore, we have opted to applying single-mode coherent states in the probing step. Their implementation is not so complex, as compared with squeezed states, 
and one can also use standard optical devices to manipulate the phase and amplitude of the coherent state. 

We are going to describe a single-mode coherent state via the basis of Fock states and the Wigner function. 
The first characterization will help us to compute the probability distribution obtained in the photodetector. 
The second characterization is important when we consider the action of the displacement operator and the description of the probability distribution in 
a heterodyne detection. 

Let $\{\ket{n}\}$, $n=0,1,2, \ldots$, be the basis of Fock states. A single-mode coherent state $\ket{\alpha}$ can be written as
\begin{equation}
  \ket{\alpha} = e^{\frac{1}{2}|\alpha|^2}\sum_{n=0}^\infty \frac{\alpha^n}{\sqrt{n!}}\ket{n}.
\end{equation}Its Wigner function reads
\begin{equation}
  W_\alpha(\mathbf{r}) = \frac{2}{\pi} e^{-(\mathbf{r} - \mathbf{\overline{r}})^T(\mathbf{r} - \mathbf{\overline{r}})},
\end{equation}where
\begin{equation}
  \mathbf{\overline{r}} = 
  \begin{pmatrix}
    \sqrt{2}\text{Re}\{\alpha\}\\
    \sqrt{2}\text{Im}\{\alpha\}
  \end{pmatrix}.
\end{equation}Since the Wigner function gives a quasiprobability description of a quantum state, we see that a 
coherent state is Gaussian.

\subsection{Encoding and Decoding Protocols}
\label{subsec:Encoding}

The encoding process consists in storing the information to be read in the future. Suppose we have a information vector $\mathbf{i}$ to be stored.
The noise due to imperfect storing or reading of the information vector ininherently prohibit us to read the information perfectly. 
We need to add redundancies in order to overcome this issue, and, in our case, using classical codes. So, we firstly produce a codeword over 
$\mathbb{F}_2$ to be written in the memory cell labels. 
Consider the information vector is a $K$ bits string, the encoder uses $\mathbf{i}$ 
to derive the codeword $\mathbf{c} = (c_1,\ldots,c_N)$ of length $N$. If the classical code is constructed over $\mathbb{F}_{2^s}$, with $s>1$ an integer, 
then it is needed to choose a basis of $\mathbb{F}_{2^s}$ over $\mathbb{F}_{2}$ and represent each coordinate $c_j$, $1\leq j\leq N$, in that basis. 
This basis expansion process will be implemented when we use Reed-Solomon codes.
After producing the codeword $\mathbf{c}$ over $\mathbb{F}_2$, we can move to the second step, which consists of associating each coordinate of $\mathbf{c}$ 
to the corresponding quantum memory cell index.

Observe that we are assuming that the quantum memory cell are composed by two quantum channels. However, the same method can be extended to quantum memory cells 
with cardinality $p^s$, where $p$ is a prime number and $s\geq 1$ is an integer.

The decoding protocol consists of two parts. The first one is measuring the probing state in order to estimate the quantum memory cell. 
After probing the set of $N$ memory cells, a noisy string or vector is obtained.
Below we present in details the measurement apparatus used through the paper, the heterodyne and Dolinar receiver. After producing the noisy vector, a decoding 
algorithm computes the best candidate of codeword for the classical code in consideration. The decoding algorithms used for the codes used in this paper have been 
discussed in Section~\ref{sec:Preliminaries}.

\subsubsection{Heterodyne Receiver with Maximum Likelihood Estimator}

Suppose we plan to recover the information stored in the memory cells, then we probe each memory cell with a coherent state. 
The output state contains the information stored in the memory cell. The lossy channel will give $\ket{\kappa_x \alpha}$, for $x=0,1$, 
once applied to the coherent state $\ket{\alpha}$. So, we need to retrieve the information about $x$.
For this goal, the heterodyne and Dolinar receiver are used. The heterodyne receiver is presented
in this subsection and Dolinar receiver in the following. 

Heterodyne receivers are used when one wishes to retrieve the information stored in the parameter $\alpha$ of the coherent state~\cite{Cariolaro2015}. The POVM describing 
this receiver is
\begin{equation}
  \Pi_\beta = \frac{1}{\pi}\ketbra{\beta}{\beta}.
\end{equation}So, the probability distribution obtained after probing the memory cell in the $j$-th position and using a heterodyne receiver is given by
\begin{equation}
  p_Y(\beta) = \frac{1}{\pi}\bra{\beta}\mathcal{W}^{c_j}(\ketbra{\alpha}{\alpha})\ket{\beta},
\end{equation}where $\beta\in\mathbb{C}$ and $Y$ is the random variable describing the output probability distribution. 
After this step, we need to divide the complex plane in order to estimate if $c_j$ is equal to zero or one. This is implemented via the 
maximum likelihood estimator. The partition of the complex plane and the decision rules are described by
\begin{equation}
  \hat{c}_j = 
  \begin{cases}
    1, & \text{if }\Lambda(\beta) \geq \eta,\\
    0, & \text{if }\Lambda(\beta) < \eta,
  \end{cases}
\end{equation}where 
\begin{equation}
  \Lambda(\beta) = \frac{p_{Y|X}(\beta|1)}{p_{Y|X}(\beta|0)}
\end{equation}and
\begin{equation}
  \eta = \frac{p_0}{p_1},
\end{equation}with $p_0$ as the probability of $0$ in a codeword $\mathbf{c}$ and $p_1 = 1 - p_0$. From the Wigner function of the coherent state, we see that 
the distribution of $p_{Y|X}(\beta|x)$, $x=0,1$, is Gaussian. Therefore, the partition of the plane can be made from the parameter of the coherent state and 
the reflexivities of the memory cells.

\begin{figure}[h!]
\center
  \resizebox{.28\hsize}{!}{    
      \input{Heterodyne_receiver_schematic.tikz}
  }
\caption{Heterodyne receiver schematic.}
\end{figure}
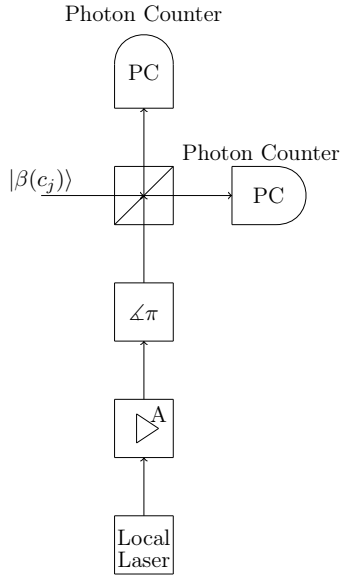

\subsubsection{Dolinar Receiver}
\label{Dolinar_receiver}

Dolinar receiver takes a different road than the heterodyne receiver~\cite{Cariolaro2015}. It is an adaptive receiver. Suppose there are two coherent states, 
$\ket{\alpha}$ and $\ket{\gamma}$, we want to distinguish. Initially, Dolinar receiver make a guess of the true state and subtract it from the state to 
be determined. This leads to the state $\ket{\psi - \phi}$, where $\psi$ is the state to be determined and $\phi$ is the guess. Next, the resulting state 
is measured in a photodetector. For the computational simulation, we use 
\begin{equation}
  Q_1 = \sum_{n=1}^\infty (1-\eta)^n\ketbra{n}{n},
\end{equation}where $1 - \eta$ is the the detection efficiency, as the click operator $Q_1$ and $Q_0 = \mathbb{I}-Q_1$ as the no-click operator. 
If there is a click, then the guess is changed and the process is repeated with the same state to be determined. Otherwise, 
the receiver declares that the true state is the one it has guessed. An illustrative schematic is shown in Figure~\ref{Dolinar_receiver}.

\begin{figure}[h!]
\center
  \resizebox{.7\hsize}{!}{    
      \input{Dolinar_receiver_schematic.tikz}
  }
\caption{Dolinar receiver schematic.}
\label{Dolinar_receiver}
\end{figure}
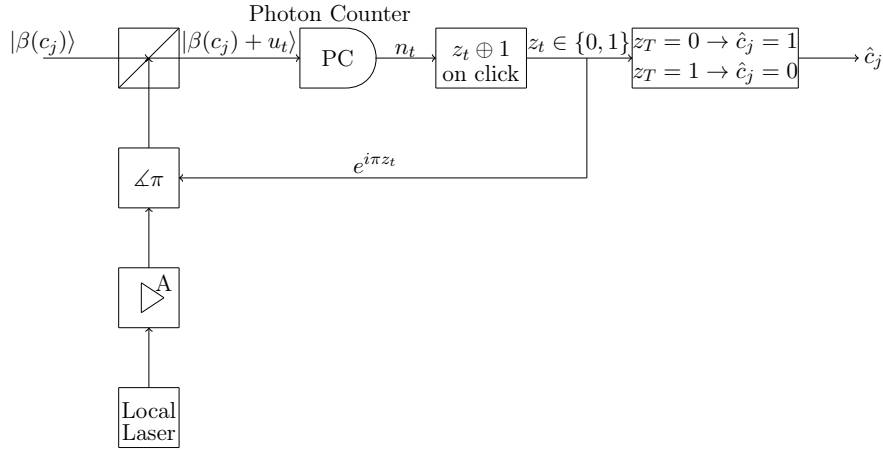

For numerical and analytical analysis of the above process, we need to clarify the feedback process showed. First of all, we are considering that 
the feedback is fast. Consider that the feedback runs $l$ times and the incoming state is $\ket{\alpha}$. Then in each round, the receiver 
has $\ket{\frac{\alpha}{\sqrt{l}}}$ as input state. This is the state that the receiver has to guess. Observe that increasing too much the number of rounds, $l$, 
in the receiver will make it more susceptible to errors. In particular, we will use $l=2$ in the following.

The second point is about the subtracted state $\ket{\phi}$. To optimize the error probability, $\ket{\phi}$ needs to be time-varying amplitude as shown in~\cite{Cariolaro2015}. 
For simplicity, we are assuming that $\ket{\phi}$ can be instantaneously changed into one of the two possible states. Even though not optimal, 
this strategy gives significant improvements when compared with the heterodyne receiver. See the following section.



%% file: Heterodyne_receiver_schematic.tikz
\begin{tikzpicture}
	\begin{pgfonlayer}{nodelayer}
		\node [style=none] (0) at (-25.75, 6) {};
		\node [style=none] (1) at (-23.75, 6) {};
		\node [style=none] (2) at (-25.75, 4) {};
		\node [style=none] (3) at (-23.75, 4) {};
		\node [style=none] (4) at (-24.75, 6) {};
		\node [style=none] (7) at (-24.75, 8) {};
		\node [style=none] (8) at (-25.75, 8) {};
		\node [style=none] (9) at (-23.75, 8) {};
		\node [style=none] (10) at (-23.75, 10) {};
		\node [style=none] (11) at (-25.75, 10) {};
		\node [style=none] (12) at (-24.75, 10) {};
		\node [style=none] (13) at (-24.75, 13) {};
		\node [style=none] (14) at (-25.75, 12) {};
		\node [style=none] (15) at (-23.75, 12) {};
		\node [style=none] (16) at (-23.75, 14) {};
		\node [style=none] (17) at (-25.75, 14) {};
		\node [style=none] (18) at (-24.75, 2) {};
		\node [style=none] (19) at (-24.75, 4) {};
		\node [style=none] (20) at (-25.75, 2) {};
		\node [style=none] (21) at (-23.75, 2) {};
		\node [style=none] (22) at (-23.75, 0) {};
		\node [style=none] (23) at (-25.75, 0) {};
		\node [style=none] (26) at (-24.75, 13) {};
		\node [style=none] (27) at (-21.75, 13) {};
		\node [style=none] (28) at (-21.75, 14) {};
		\node [style=none] (29) at (-21.75, 12) {};
		\node [style=none] (30) at (-20.25, 14) {};
		\node [style=none] (31) at (-20.25, 12) {};
		\node [style=none] (32) at (-19.25, 13) {};
		\node [style=none] (45) at (-28.25, 13) {};
		\node [style=none] (48) at (-20.5, 13) {PC};
		\node [style=none] (49) at (-20.75, 14.5) {Photon Counter};
		\node [style=none] (51) at (-28.25, 13.5) {$|\beta(c_j)\rangle$};
		\node [style=none] (54) at (-24.75, 1.25) {Local};
		\node [style=none] (56) at (-24.75, 0.5) {Laser};
		\node [style=none] (59) at (-23.75, 9) {};
		\node [style=none] (68) at (-24.75, 9) {$\measuredangle \pi$};
		\node [style=none] (70) at (-24.25, 5.5) {A};
		\node [style=none] (71) at (-25, 5.5) {};
		\node [style=none] (72) at (-25, 4.5) {};
		\node [style=none] (73) at (-24.25, 5) {};
		\node [style=none] (75) at (-24.75, 16) {};
		\node [style=none] (76) at (-25.75, 16) {};
		\node [style=none] (77) at (-23.75, 16) {};
		\node [style=none] (78) at (-25.75, 17.5) {};
		\node [style=none] (79) at (-23.75, 17.5) {};
		\node [style=none] (80) at (-24.75, 18.5) {};
		\node [style=none] (81) at (-24.75, 17.25) {PC};
		\node [style=none] (82) at (-24.75, 19.25) {Photon Counter};
	\end{pgfonlayer}
	\begin{pgfonlayer}{edgelayer}
		\draw (0.center) to (1.center);
		\draw (1.center) to (3.center);
		\draw (3.center) to (2.center);
		\draw (2.center) to (0.center);
		\draw [style=edge right arrow] (4.center) to (7.center);
		\draw (11.center) to (10.center);
		\draw (10.center) to (9.center);
		\draw (9.center) to (8.center);
		\draw (8.center) to (11.center);
		\draw [style=edge right arrow] (12.center) to (13.center);
		\draw (14.center) to (15.center);
		\draw (15.center) to (16.center);
		\draw (16.center) to (17.center);
		\draw (17.center) to (14.center);
		\draw [style=edge right arrow] (18.center) to (19.center);
		\draw (20.center) to (21.center);
		\draw (21.center) to (22.center);
		\draw (22.center) to (23.center);
		\draw (23.center) to (20.center);
		\draw [style=edge right arrow] (26.center) to (27.center);
		\draw (28.center) to (29.center);
		\draw (28.center) to (30.center);
		\draw (29.center) to (31.center);
		\draw [bend left=90, looseness=1.75] (30.center) to (31.center);
		\draw [style=edge right arrow] (45.center) to (26.center);
		\draw (71.center) to (72.center);
		\draw (72.center) to (73.center);
		\draw (73.center) to (71.center);
		\draw (14.center) to (26.center);
		\draw (26.center) to (16.center);
		\draw [style=edge right arrow] (26.center) to (75.center);
		\draw (76.center) to (77.center);
		\draw (76.center) to (78.center);
		\draw (77.center) to (79.center);
		\draw [bend left=90, looseness=1.75] (78.center) to (79.center);
	\end{pgfonlayer}
\end{tikzpicture}

%% file: Dolinar_receiver_schematic.tikz
\begin{tikzpicture}
	\begin{pgfonlayer}{nodelayer}
		\node [style=none] (0) at (-14, 6.75) {};
		\node [style=none] (1) at (-12, 6.75) {};
		\node [style=none] (2) at (-14, 4.75) {};
		\node [style=none] (3) at (-12, 4.75) {};
		\node [style=none] (4) at (-13, 6.75) {};
		\node [style=none] (7) at (-13, 8.75) {};
		\node [style=none] (8) at (-14, 8.75) {};
		\node [style=none] (9) at (-12, 8.75) {};
		\node [style=none] (10) at (-12, 10.75) {};
		\node [style=none] (11) at (-14, 10.75) {};
		\node [style=none] (12) at (-13, 10.75) {};
		\node [style=none] (13) at (-13, 13.75) {};
		\node [style=none] (14) at (-14, 12.75) {};
		\node [style=none] (15) at (-12, 12.75) {};
		\node [style=none] (16) at (-12, 14.75) {};
		\node [style=none] (17) at (-14, 14.75) {};
		\node [style=none] (18) at (-13, 2.75) {};
		\node [style=none] (19) at (-13, 4.75) {};
		\node [style=none] (20) at (-14, 2.75) {};
		\node [style=none] (21) at (-12, 2.75) {};
		\node [style=none] (22) at (-12, 0.75) {};
		\node [style=none] (23) at (-14, 0.75) {};
		\node [style=none] (26) at (-13, 13.75) {};
		\node [style=none] (27) at (-8, 13.75) {};
		\node [style=none] (28) at (-8, 14.75) {};
		\node [style=none] (29) at (-8, 12.75) {};
		\node [style=none] (30) at (-6.5, 14.75) {};
		\node [style=none] (31) at (-6.5, 12.75) {};
		\node [style=none] (32) at (-5.5, 13.75) {};
		\node [style=none] (33) at (-3.5, 13.75) {};
		\node [style=none] (34) at (-3.5, 14.75) {};
		\node [style=none] (35) at (-3.5, 12.75) {};
		\node [style=none] (36) at (-0.5, 12.75) {};
		\node [style=none] (37) at (-0.5, 14.75) {};
		\node [style=none] (38) at (-0.5, 13.75) {};
		\node [style=none] (39) at (3, 13.75) {};
		\node [style=none] (40) at (3, 14.75) {};
		\node [style=none] (41) at (3, 12.75) {};
		\node [style=none] (42) at (8.5, 12.75) {};
		\node [style=none] (43) at (8.5, 14.75) {};
		\node [style=none] (44) at (10.5, 13.75) {};
		\node [style=none] (45) at (-16.5, 13.75) {};
		\node [style=none] (46) at (8.5, 13.75) {};
		\node [style=none] (48) at (-6.75, 13.75) {PC};
		\node [style=none] (49) at (-7, 15.25) {Photon Counter};
		\node [style=none] (51) at (-16.5, 14.25) {$|\beta(c_j)\rangle$};
		\node [style=none] (54) at (-13, 2) {Local};
		\node [style=none] (56) at (-13, 1.25) {Laser};
		\node [style=none] (57) at (1.5, 13.75) {};
		\node [style=none] (58) at (1.5, 9.75) {};
		\node [style=none] (59) at (-12, 9.75) {};
		\node [style=none] (60) at (-4.5, 14) {$n_t$};
		\node [style=none] (61) at (-2, 14) {$z_t\oplus 1$};
		\node [style=none] (62) at (-2, 13.25) {on click};
		\node [style=none] (63) at (1.25, 14.25) {$z_t\in\{0,1\}$};
		\node [style=none] (64) at (5.75, 14.25) {$z_T = 0\rightarrow \hat{c}_j = 1$};
		\node [style=none] (65) at (5.75, 13.25) {$z_T = 1\rightarrow \hat{c}_j = 0$};
		\node [style=none] (66) at (-5.5, 10.25) {$e^{i\pi z_t}$};
		\node [style=none] (67) at (-10, 14.25) {$|\beta(c_j) + u_t\rangle$};
		\node [style=none] (68) at (-13, 9.75) {$\measuredangle \pi$};
		\node [style=none] (70) at (-12.5, 6.25) {A};
		\node [style=none] (71) at (-13.25, 6.25) {};
		\node [style=none] (72) at (-13.25, 5.25) {};
		\node [style=none] (73) at (-12.5, 5.75) {};
		\node [style=none] (74) at (11, 13.75) {$\hat{c}_j$};
	\end{pgfonlayer}
	\begin{pgfonlayer}{edgelayer}
		\draw (0.center) to (1.center);
		\draw (1.center) to (3.center);
		\draw (3.center) to (2.center);
		\draw (2.center) to (0.center);
		\draw [style=edge right arrow] (4.center) to (7.center);
		\draw (11.center) to (10.center);
		\draw (10.center) to (9.center);
		\draw (9.center) to (8.center);
		\draw (8.center) to (11.center);
		\draw [style=edge right arrow] (12.center) to (13.center);
		\draw (14.center) to (15.center);
		\draw (15.center) to (16.center);
		\draw (16.center) to (17.center);
		\draw (17.center) to (14.center);
		\draw [style=edge right arrow] (18.center) to (19.center);
		\draw (20.center) to (21.center);
		\draw (21.center) to (22.center);
		\draw (22.center) to (23.center);
		\draw (23.center) to (20.center);
		\draw [style=edge right arrow] (26.center) to (27.center);
		\draw (28.center) to (29.center);
		\draw (28.center) to (30.center);
		\draw (29.center) to (31.center);
		\draw [bend left=90, looseness=1.75] (30.center) to (31.center);
		\draw [style=edge right arrow] (32.center) to (33.center);
		\draw (34.center) to (37.center);
		\draw (37.center) to (36.center);
		\draw (36.center) to (35.center);
		\draw (35.center) to (34.center);
		\draw (40.center) to (41.center);
		\draw (41.center) to (42.center);
		\draw (42.center) to (43.center);
		\draw (43.center) to (40.center);
		\draw [style=edge right arrow] (38.center) to (39.center);
		\draw [style=edge right arrow] (45.center) to (26.center);
		\draw [style=edge right arrow] (46.center) to (44.center);
		\draw [style=edge right arrow] (58.center) to (59.center);
		\draw (57.center) to (58.center);
		\draw (71.center) to (72.center);
		\draw (72.center) to (73.center);
		\draw (73.center) to (71.center);
		\draw (14.center) to (26.center);
		\draw (26.center) to (16.center);
	\end{pgfonlayer}
\end{tikzpicture}

%% file: errorMitigation.tex
Consider memory cells $\{\mathcal{W}^0,\mathcal{W}^1\}$, where $p_0 = p_1 = 1/2$, with reflexivities $\kappa_0$ and $\kappa_2$, respectively. 
Suppose a string of memory cells are encoded via a classical code $\mathcal{C}$. The information stored in the string is retrieved by probing 
them via a coherent state. From the result obtained, we analyze the error probability between the actual information $\mathbf{i}$ encoded by $\mathcal{C}$ 
and the estimated information $\hat{\mathbf{i}}$ after decoding the noisy string after the probing step. 

\begin{figure}[h!]
    \centering
    \settototalheight{\dimen0}{
        \includegraphics[width=\columnwidth]{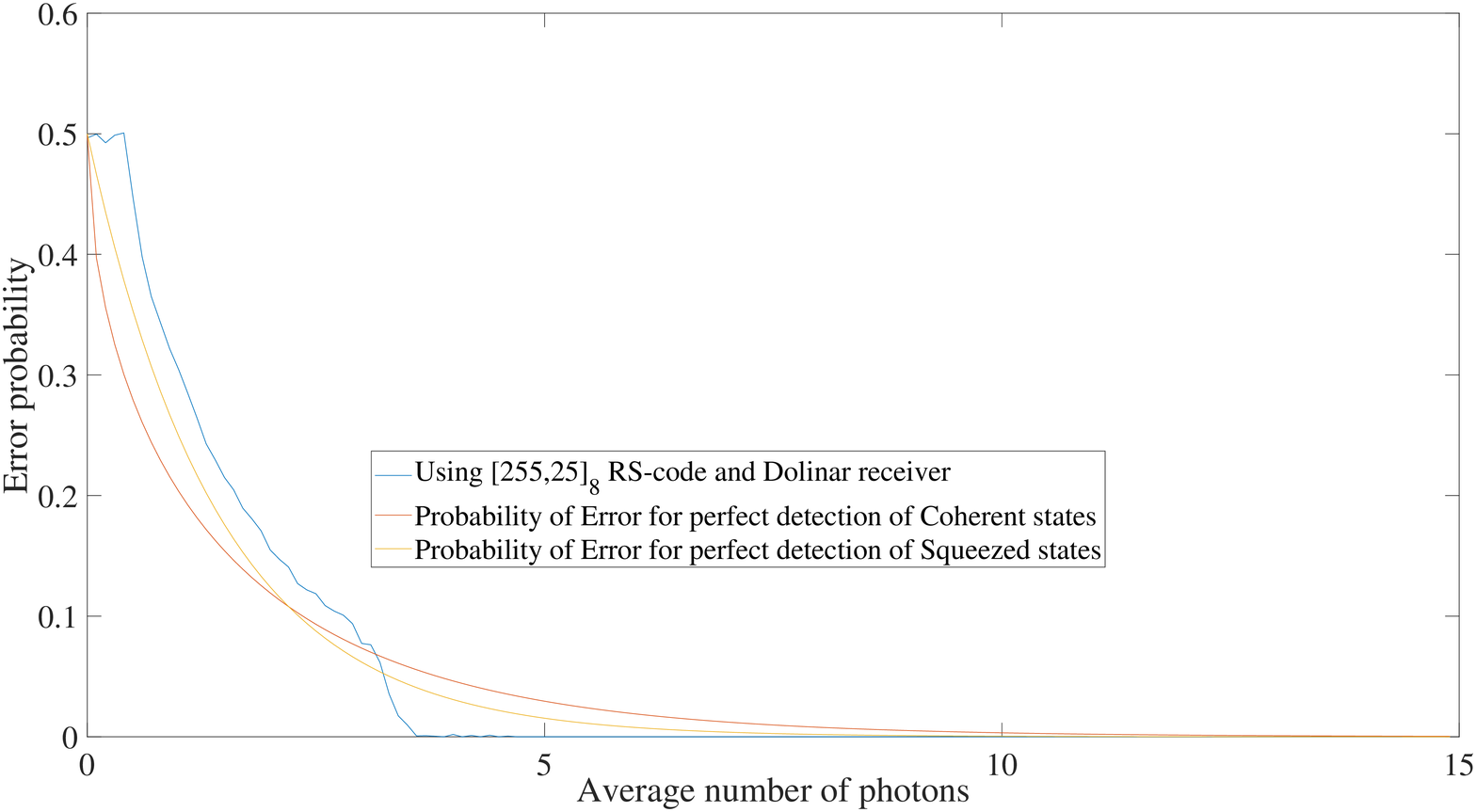}
    }%
    \includegraphics[width=\columnwidth]{n_255-k_25_Dolinar_2.eps}%
    \llap{
        \raisebox{\dimen1+4.15cm}{
            \includegraphics[height=4cm]{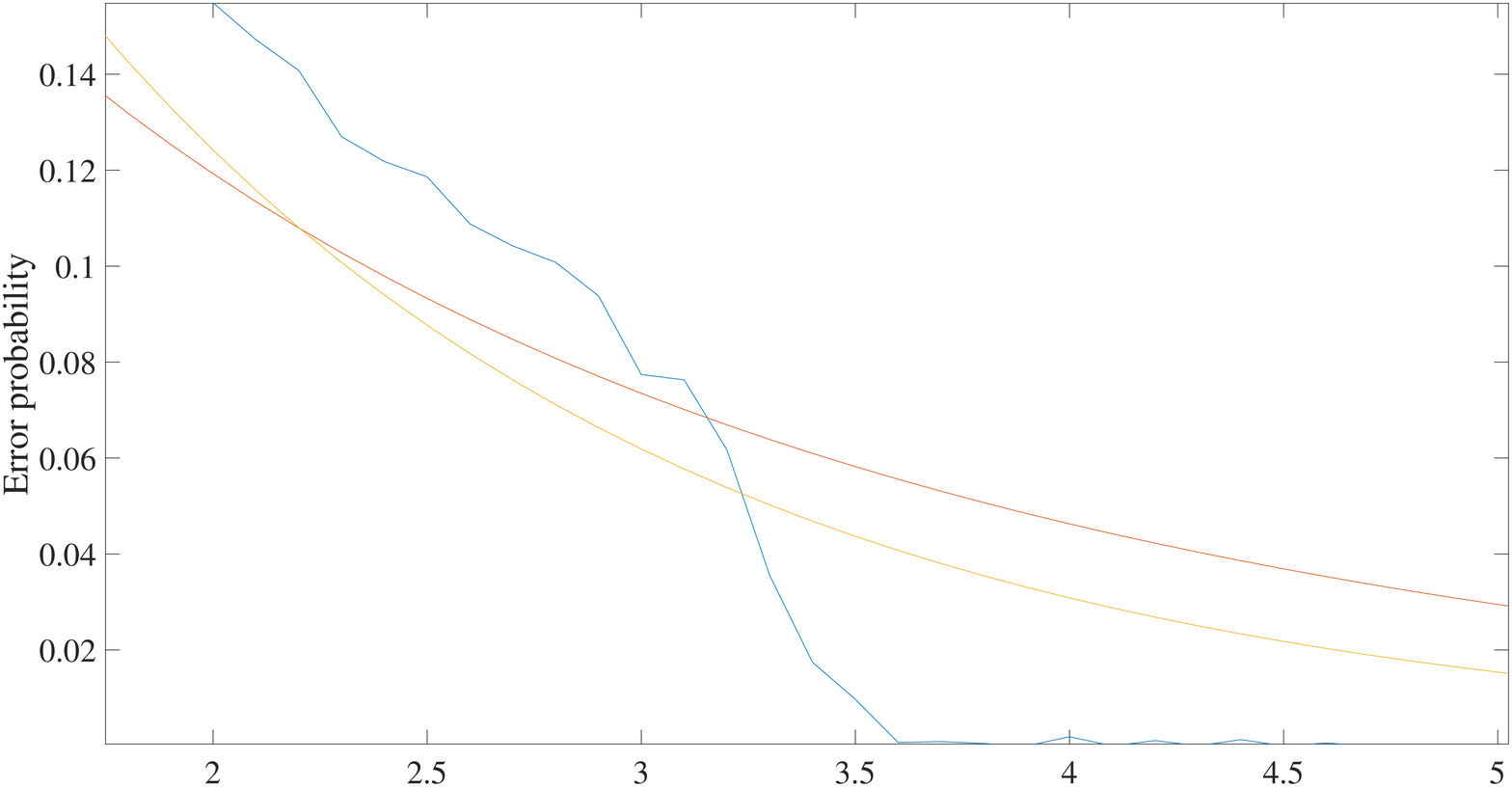}%
        }
    }
    \caption{Error probability decay using $[255,25]_8$ Reed-Solomon code and Dolinar receiver for $\kappa_0 = 0.1$ and $\kappa_1 = 0.95$.}
  \label{fig:SampleRS}
\end{figure}

A sample of plot we produce is shown in Fig.~\ref{fig:SampleRS}. For a fixed rate $R = \frac{K}{N}$, where $N$ and $K$ are the length and dimension of the 
classical code, respectively, we compute the error probability for several values of average number of photons. The receiver is also fixed. For the particular 
case of Fig.~\ref{fig:SampleRS}, we used a heterodyne receiver. The analysis described in details for the heterodyne and Dolinar receivers below focus on the 
cross point or threshold where using the classical code gives improvement compared to optimal receiver using coherent or squeezed states. As an example, 
the value of the threshold in using RS code with rate $R = 25/255$ and Dolinar receiver is equal to $3.2$, as one can see in Fig.~\ref{fig:SampleRS}. 
Notice that the threshold depends on the slope of the error probability curve for each code and receiver. 
Therefore, it depends on the family of the code, the parameters of the code, and the receiver used. 
See below the error mitigation behavior analyzed under the perspective of the threshold for three families of classical codes, with several rates, and 
two types of receivers.

\subsection{Heterodyne Receiver}

For fixed values of transmissivities, $\kappa_0$ and $\kappa_1$, the analysis given in this paper considers the behavior of error 
mitigation for different codes. Because of analytical difficulties in deriving a closed formula for the error probability using classical 
codes, we adopted numerical simulations as the aid tool for comparison between our case of study and the literature. 

\begin{figure}[h!]
  \centering
  \includegraphics[width=\linewidth]{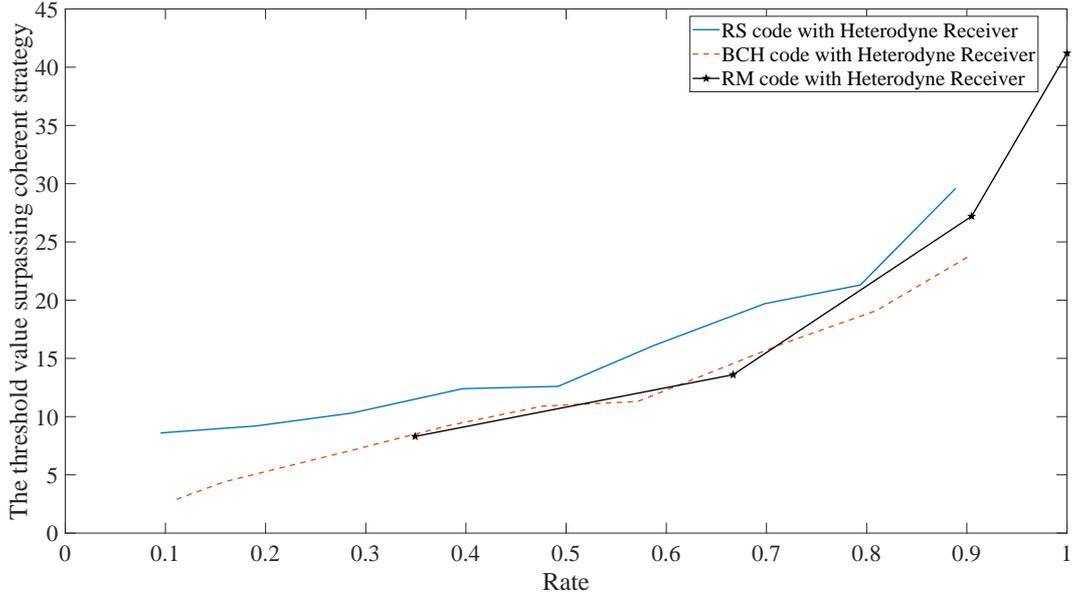}
  \caption{Error mitigation using Reed-Solomon, BCH, or Reed-Muller codes. 
           All codes are considered in conjunction with a heterodyne receiver. 
           The memory cells have transmissivities equal to $\kappa_0 = 0.1$ and $\kappa_1 = 0.95$.}
  \label{fig:RS-BCH-RM_Heterodyne_.1_.95}
\end{figure}

The first characterization is presented in Fig.~\ref{fig:RS-BCH-RM_Heterodyne_.1_.95}. The behaviors of the three codes are similar. 
First of all, as the rate increases, it also increases the value of the threshold. Increasing the rate means that we have less redundancy 
and, therefore, the code has the error-correction capability decreased, which impose the need for probing with higher energy in order to 
surpass the optimal strategies. Secondly, the threshold values go up to 45 photons. This means that we can keep in the quantum regime for 
surpassing the optimal strategies when using classical codes. 

Comparing between codes, BCH codes and Reed-Muller codes have similar performances. It is important to remember that there are some
limitations in using Reed-Muller codes since the choice of parameters has not so much freedom when compared with BCH and Reed-Solomon codes. 
For Reed-Solomon codes, we see that they request probing with more energy but such increase of energy is less than 5 photons. 
Therefore, in practical applications, one would chose the code that has encoder and decoder at disposal with the lowest complexities.

Next, the behavior of the error mitigation is investigated for different values of transmissivities and using Reed-Muller codes. 
The first aspect one can see is that the performance highly depends on the difference between the transmissivities $\kappa_0$ and $\kappa_1$.
For transmissivities where $\kappa_0$ and $\kappa_1$ are closer, the output probe states have position and momentum which are closer. Therefore,
the heterodyne receiver in conjunction with maximum likelihood estimator has a higher probability of estimating the value encoded in the memory 
cell erroneously. This will imply in a receiver output string with more errors. For a code with the same rate to surpass the optimal receiver, it 
need more probing energy so that error probability decreases. This is the reason why one need more energy for $\kappa_0 = 0.3$ and $\kappa_1 = 0.75$ 
and why the slop of the curve is higher than the case with $\kappa_0 = 0.1$ and $\kappa_1 = 0.95$. For Reed-Solomon and BCH codes, we have the same 
behavior. 

\begin{figure}[h!]
  \centering
  \includegraphics[width=\linewidth]{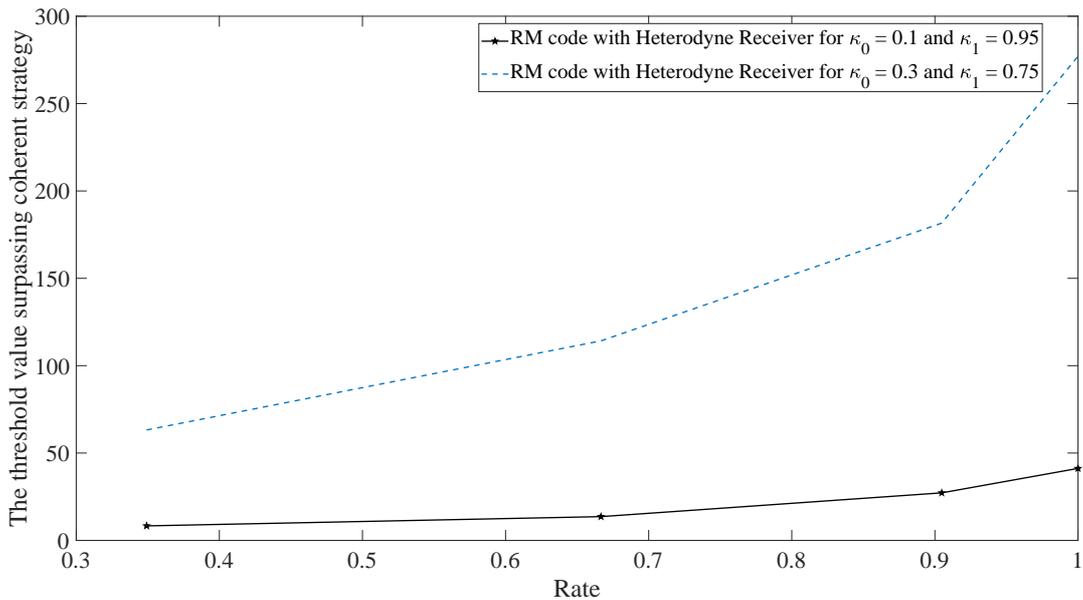}
  \caption{Comparison of performance using Reed-Muller codes for memory cells with different transmissivities. 
  Solid line refers to memory cells with trasmittivities given by $\kappa_0 = 0.1$ and $\kappa_1 = 0.95$. 
  Dotted line refers to memory cells with transmittivities $\kappa_0 = 0.3$ and $\kappa_1 = 0.75$.}
  \label{fig:Threshold_CoherentState_RM}
\end{figure}

\subsection{Dolinar Receiver}
Dolinar receiver is a powerful receiver that can be used for achieving the Helstrom bound on the error probability of distinguishing 
quantum states. As mentioned before, it also comes with a inherit complexity due to its adaptive character. Even though, it can be implement
in practice~\cite{Cook2007}. For the analysis shown below, we need to impose some defects to the Dolinar receiver, otherwise the results obtained would be 
unrealistic. The structure and procedure for implementing the Dolinar receiver is the one explained in Section~\ref{Dolinar_receiver}. 
The defect imposed for the results below is over the efficiency of the photodetector, which we consider to be equal to $0.9$. This value 
is close to the one obtained with the current technology for some photodetectors. One additional defect that could be added is dark counting. 
However, to consider an experiment with dark counting would need additional parameters, such as rate of measurement over the memory cells. 
The complexity in dealing with these details could fade the importance of the codes used and mislead the analysis. 

\begin{figure*}[h!]
  \centering
  \subfloat[\label{1a}]{%
      \includegraphics[width=0.495\linewidth]{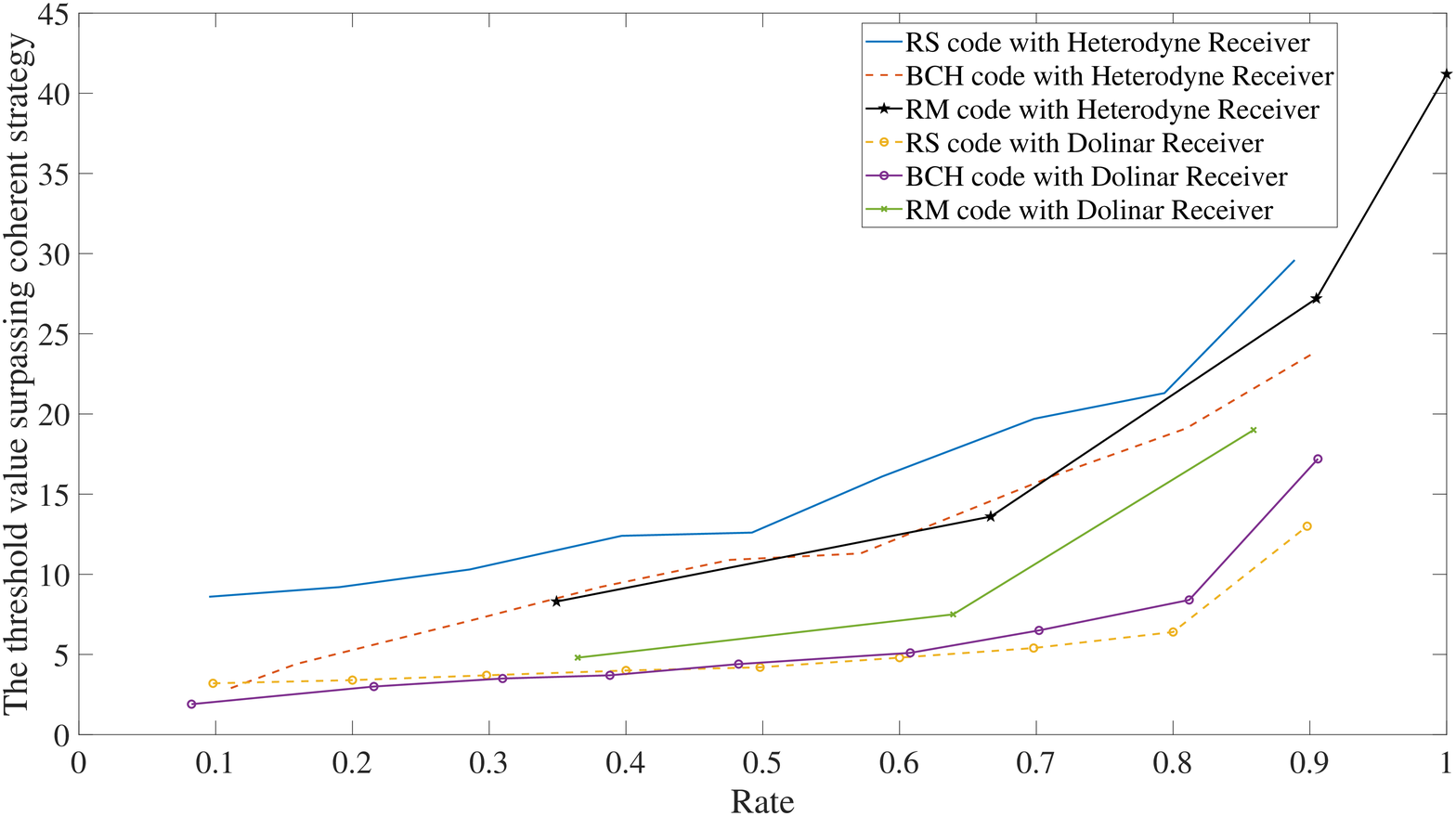}}
  \hfill
  \subfloat[\label{1b}]{%
      \includegraphics[width=0.495\linewidth]{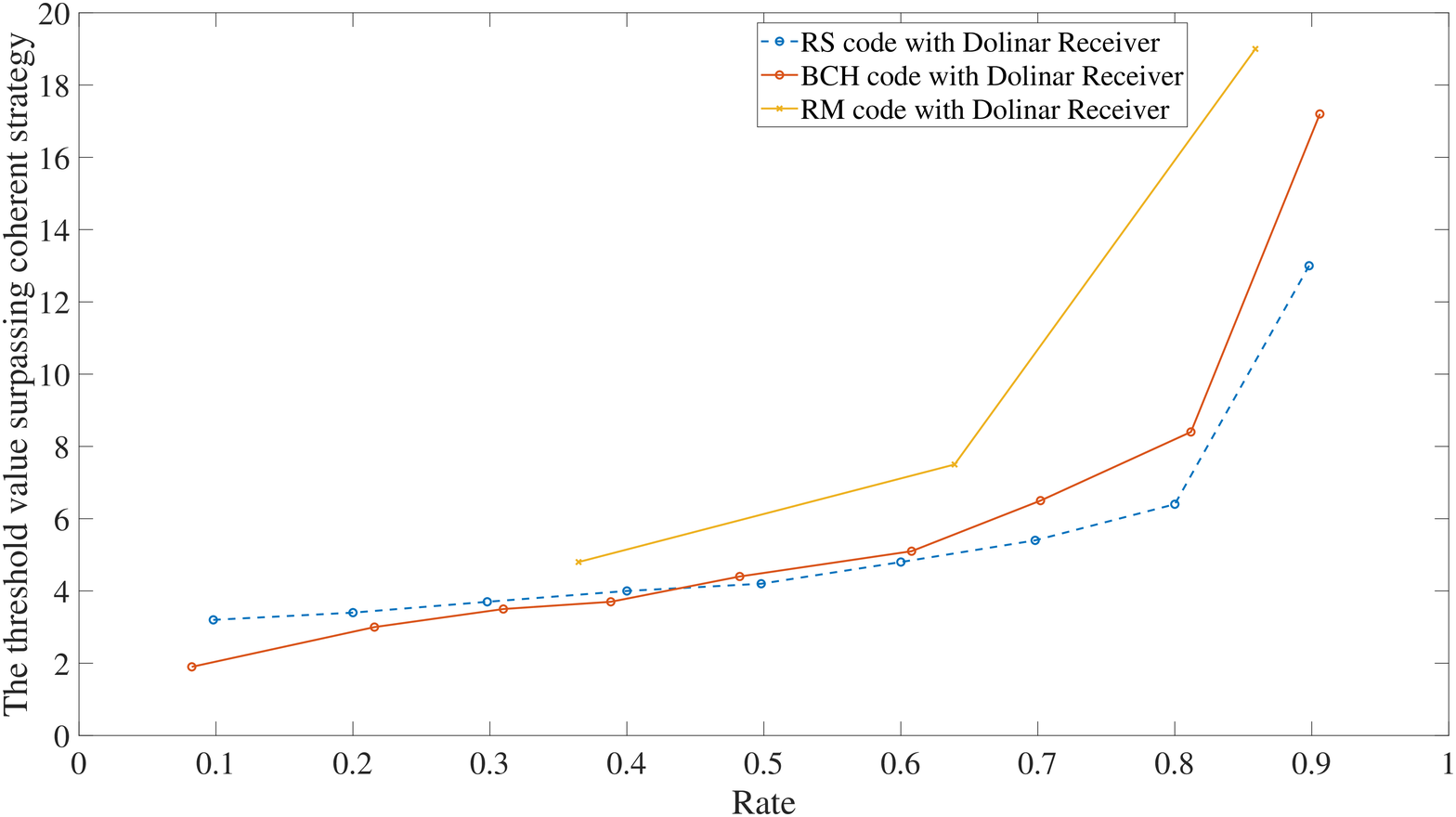}
  }
  \caption{Comparison between heterodyne receiver (a) and Dolinar receiver (b) when using Reed-Solomon, BCH, and Reed-Muller 
    codes. We are considering that the photodetector used in the Dolinar receiver has efficiency equal to $0.9$ and 
    memory cells have transmissivities $\kappa_0 = 0.1$ and $\kappa_1 = 0.95$. 
    Figure (a) displays the three families of codes with heterodyne and Dolinar receiver.
  }
  \label{fig:RS-BCH-RM_Heterodyne_VERSUS_Dolinar}
\end{figure*}

Let the reflexivities be $\kappa_0 = 0.1$ and $\kappa_1 = 0.95$. It is shown in Fig.~\ref{fig:RS-BCH-RM_Heterodyne_VERSUS_Dolinar} the performance 
using Reed-Solomon, BCH, or Reed-Muller codes in conjunction to a heterodyne or Dolinar receiver. Reed-Solomon and BCH codes in conjunction to Dolinar 
receiver give the best performance between them all. For low rates, BCH codes with heterodyne or Dolinar receiver give similar results. In particular, 
for rates below $0.1$, the threshold value of using BCH codes with heterodyne receiver is 3.2 and with Dolinar receiver is 1.9. However, the difference 
between heterodyne and Dolinar receiver accentuates for higher rate. 

The performance of using Reed-Solomon or BCH codes in conjunction with Dolinar receiver is quite similar. For some rate $R\in[0.4,0.5]$ we have that 
Reed-Solomon codes show better error mitigation than BCH codes. However, for practical reasons, BCH codes may be the best choice. The reason is that 
Reed-Solomon codes are constructed in an extended field $\mathbb{F}_{2^s}$, for some integer $s$, of $\mathbb{F}_2$. This imposes the need of 
constantly having to decompose the coordinates of the codewords in a basis of $\mathbb{F}_{2^s}$ over $\mathbb{F}_2$. Such decomposition is implemented 
in the encoding and when writing the received string to feed the decoder. It is a not so demanding process but may be substantial for choosing in favor of
BCH codes.

There a last point to consider. The accomplishments presented using Dolinar receiver could be due to the receiver and, therefore, no code is needed. 
However, this is not true. Since the efficiency of the photodetector is below one, we have that the error probability without code but using Dolinar 
receiver is bounded from below. In particular, for the efficiency of $0.9$ considered in our simulation and for any value of average number of photons, 
the error probability is always above $0.12\%$. So, to improve further the error mitigation, one needs to amend additional tools, such as classical codes.

%% file: conclusion.tex
We have shown that classical error-correcting codes can be used as a tool to reduce error probability in quantum reading. 
They have been applied in a short length range.
Even so, error mitigation is accomplished. Above all they can 
exhibit improvements when compared with optimal strategies using coherent or two-mode squeezed states once a threshold is crossed. 
All of these for two types of receivers, heterodyne receiver or Dolinar receiver. We also studied the situation when the channel transmissivities 
are closer. It was shown that the error mitigation deteriorates but one can still surpass the optimal strategies probing with a state with higher 
energy. The same conclusion is obtained for the Dolinar receiver. 
As an overall conclusion of the analysis presented, BCH codes are the optimal code choice for error mitigation in the quantum reading task 
among the families of codes considered.

Several aspects could be further investigated in future works. Firstly, how the error probability mitigation behaves using codes with larger lengths and different rates. 
Some example of possible choice of codes are LDPC and Turbo codes.
It is expected that using codes with larger lengths and lower rates reduce the error probability. However, we do not know how this 
reduction will behave in this or in different scenarios. Secondly, analytical upper and lower bounds need 
to be obtained for a broader understanding of the current results. 
Lastly, it is expected that classical codes performance depends on the weight distribution of the code. 
This result may help to derive the error bounds mentioned before.

%% file: _2021_ArXiv-Francisco_Stefano-Error_Mitigation_using_Classical_Codes.bbl
\begin{thebibliography}{10}
\providecommand{\url}[1]{#1}
\csname url@samestyle\endcsname
\providecommand{\newblock}{\relax}
\providecommand{\bibinfo}[2]{#2}
\providecommand{\BIBentrySTDinterwordspacing}{\spaceskip=0pt\relax}
\providecommand{\BIBentryALTinterwordstretchfactor}{4}
\providecommand{\BIBentryALTinterwordspacing}{\spaceskip=\fontdimen2\font plus
\BIBentryALTinterwordstretchfactor\fontdimen3\font minus
  \fontdimen4\font\relax}
\providecommand{\BIBforeignlanguage}[2]{{%
\expandafter\ifx\csname l@#1\endcsname\relax
\typeout{** WARNING: IEEEtran.bst: No hyphenation pattern has been}%
\typeout{** loaded for the language `#1'. Using the pattern for}%
\typeout{** the default language instead.}%
\else
\language=\csname l@#1\endcsname
\fi
#2}}
\providecommand{\BIBdecl}{\relax}
\BIBdecl

\bibitem{Hirvensalo2003}
M.~Hirvensalo, \emph{Quantum Computing}.\hskip 1em plus 0.5em minus 0.4em\relax
  Springer Berlin Heidelberg, 2003.

\bibitem{Cariolaro2015}
G.~Cariolaro, \emph{Quantum Communications}.\hskip 1em plus 0.5em minus
  0.4em\relax Springer-Verlag GmbH, 2015.

\bibitem{AndreasWinter2020}
S.~Mancini and A.~Winter, \emph{A Quantum Leap in Information Theory}.\hskip
  1em plus 0.5em minus 0.4em\relax WSPC, 2020.

\bibitem{Childs2000}
A.~M. Childs, J.~Preskill, and J.~Renes, ``Quantum information and precision
  measurement,'' \emph{Journal of Modern Optics}, vol.~47, no. 2-3, pp.
  155--176, feb 2000.

\bibitem{Acin2001}
A.~Ac{\'{\i}}n, ``Statistical distinguishability between unitary operations,''
  \emph{Physical Review Letters}, vol.~87, no.~17, oct 2001.

\bibitem{Gilchrist2005}
A.~Gilchrist, N.~K. Langford, and M.~A. Nielsen, ``Distance measures to compare
  real and ideal quantum processes,'' \emph{Physical Review A}, vol.~71, no.~6,
  jun 2005.

\bibitem{Sacchi2005}
M.~F. Sacchi, ``Optimal discrimination of quantum operations,'' \emph{Physical
  Review A}, vol.~71, no.~6, jun 2005.

\bibitem{Sacchi2005a}
------, ``Entanglement can enhance the distinguishability of
  entanglement-breaking channels,'' \emph{Physical Review A}, vol.~72, no.~1,
  jul 2005.

\bibitem{Wang2006}
G.~Wang and M.~Ying, ``Unambiguous discrimination among quantum operations,''
  \emph{Physical Review A}, vol.~73, no.~4, apr 2006.

\bibitem{Duan2009}
R.~Duan, Y.~Feng, and M.~Ying, ``Perfect distinguishability of quantum
  operations,'' \emph{Physical Review Letters}, vol. 103, no.~21, nov 2009.

\bibitem{Hayashi2009}
M.~Hayashi, ``Discrimination of two channels by adaptive methods and its
  application to quantum system,'' \emph{{IEEE} Transactions on Information
  Theory}, vol.~55, no.~8, pp. 3807--3820, aug 2009.

\bibitem{Harrow2010}
A.~W. Harrow, A.~Hassidim, D.~W. Leung, and J.~Watrous, ``Adaptive versus
  nonadaptive strategies for quantum channel discrimination,'' \emph{Physical
  Review A}, vol.~81, no.~3, mar 2010.

\bibitem{Rexiti2021}
M.~Rexiti and S.~Mancini, ``Discriminating qubit amplitude damping channels,''
  \emph{Journal of Physics A: Mathematical and Theoretical}, vol.~54, no.~16,
  p. 165303, mar 2021.

\bibitem{Pirandola2011}
S.~Pirandola, ``Quantum reading of a classical digital memory,'' \emph{Physical
  Review Letters}, vol. 106, no.~9, mar 2011.

\bibitem{Pirandola2011a}
S.~Pirandola, C.~Lupo, V.~Giovannetti, S.~Mancini, and S.~L. Braunstein,
  ``Quantum reading capacity,'' \emph{New Journal of Physics}, vol.~13, no.~11,
  p. 113012, nov 2011.

\bibitem{Das2019}
S.~Das and M.~M. Wilde, ``Quantum rebound capacity,'' \emph{Physical Review A},
  vol. 100, no.~3, sep 2019.

\bibitem{Pereira2020}
F.~R.~F. Pereira and S.~Mancini, ``Polar codes for quantum reading,''
  arXiv:2012.07198.

\bibitem{Pereira2021}
------, ``Stabilizer codes for open quantum systems,'' arXiv:2107.11914.

\bibitem{Pellikaan2017}
R.~Pellikaan, X.-W. Wu, S.~Bulygin, and R.~Jurrius, \emph{Codes, Cryptology and
  Curves with Computer Algebra}.\hskip 1em plus 0.5em minus 0.4em\relax
  Cambridge University Press, oct 2017.

\bibitem{Cook2007}
R.~L. Cook, P.~J. Martin, and J.~M. Geremia, ``Optical coherent state
  discrimination using a closed-loop quantum measurement,'' \emph{Nature}, vol.
  446, no. 7137, pp. 774--777, apr 2007.

\bibitem{Banchi2020}
L.~Banchi, Q.~Zhuang, and S.~Pirandola, ``Quantum-enhanced barcode decoding and
  pattern recognition,'' \emph{Physical Review Applied}, vol.~14, no.~6, dec
  2020.

\bibitem{HuffmanVera:Book}
W.~C. Huffman and V.~Pless, \emph{Fundamentals of Error-Correcting
  Codes}.\hskip 1em plus 0.5em minus 0.4em\relax Cambridge University Press,
  jun 2003.

\bibitem{Reed1954}
I.~Reed, ``A class of multiple-error-correcting codes and the decoding
  scheme,'' \emph{Transactions of the IRE Professional Group on Information
  Theory}, vol.~4, no.~4, pp. 38--49, sep 1954.

\end{thebibliography}
